\documentclass[twocolumn,showpacs,preprintnumbers,prl,aps,amssymb,superscriptaddress]{revtex4}
\usepackage{graphicx}
\usepackage{dcolumn}
\usepackage{bm}

\begin{document}

\title{Spectroscopic Evidence for Multiple Order Parameter Components in the Heavy Fermion Superconductor CeCoIn$_5$}

\author{P.~M.~C.~Rourke}
\affiliation{Department of Physics, University of Toronto, Toronto,
Ontario, M5S1A7 Canada}
\author{M.~A.~Tanatar}
\altaffiliation{Permanent address: Institute of Surface Chemistry,
N.A.S. Ukraine, Kyiv, Ukraine.}
\affiliation{Department of Physics,
University of Toronto, Toronto, Ontario, M5S1A7 Canada}
\author{C.~S.~Turel}
\affiliation{Department of Physics, University of Toronto, Toronto,
Ontario, M5S1A7 Canada}
\author{J.~Berdeklis}
\affiliation{Department of Physics, University of Toronto, Toronto,
Ontario, M5S1A7 Canada}
\author{C.~Petrovic}
\affiliation{Department of Physics, Brookhaven National Laboratory,
Upton, New York 11973, USA}
\author{J.~Y.~T.~Wei}
\affiliation{Department of Physics, University of Toronto, Toronto,
Ontario, M5S1A7 Canada}

\date{\today}

\begin{abstract}

Point-contact spectroscopy was performed on single crystals of the
heavy-fermion superconductor CeCoIn$_5$ between 150 mK to 2.5 K. A
pulsed measurement technique ensured minimal Joule heating over a
wide voltage range. The spectra show Andreev-reflection
characteristics with \emph{multiple} structures which depend on
junction impedance. Spectral analysis using the generalized
Blonder-Tinkham-Klapwijk formalism for \emph{d}-wave pairing
revealed two \emph{coexisting} order parameter components with
amplitudes $\Delta_1$ = 0.95 $\pm$ 0.15 meV and $\Delta_2$ = 2.4
$\pm$ 0.3 meV, which evolve differently with temperature. These
observations indicate a highly unconventional pairing mechanism,
possibly involving multiple bands.
\end{abstract}

\pacs{74.70.Tx, 74.50+r, 74.20.Rp, 74.45.+c} \maketitle

The discovery of the heavy-fermion superconductor CeCoIn$_5$ has
attracted widespread interest in the field of superconductivity
\cite{Petrovic}. Besides having the highest critical temperature
$T_c=2.3$ K among heavy-fermion materials, CeCoIn$_5$ also shares
some unconventional properties with the high-$T_c$ cuprates. First,
CeCoIn$_5$ has shown pronounced non-Fermi liquid behaviors,
suggestive of quantum critical phenomena that could arise from
competing orders~\cite{Sidorov,Paglione}. Second, CeCoIn$_5$ has
shown low-energy quasiparticle excitations and a power-law
temperature dependence in the NMR spin relaxation, indicative of
nodes in the superconducting energy gap
\cite{Movshovich-unconv,NQR,Izawa-aniz,Aoki}. These nodal
charateristics are consistent with $d$-wave pairing symmetry
\cite{TsueiKirtleyrevu}, which could be produced by
antiferromagnetic fluctuations~\cite{MonthouxLonzarich}. Unlike the
cuprates, on the other hand, CeCoIn$_5$ is an intermetallic compound
with multiple sheets on the Fermi surface~\cite{Bandstructure,dHvA}.
Such complex Fermi topology could involve several bands in the
pairing process, giving rise to multiple pair
potentials~\cite{Suhl,Mazin,MgB2}.

Point-contact spectroscopy (PCS) has been a proven microscopic
technique for studying unconventional superconductors. For the
high-$T_c$ cuprates, PCS provided the earliest measurements of the
superconducting gap spectra~\cite{Zasadzinsky}. In MgB$_2$, PCS was
key in revealing two coexisting \emph{s}-wave gaps~\cite{Szabo}. PCS
has been previously performed on several heavy-fermion
superconductors~\cite{Nowack,HasselbachKirtley,DeWilde,HFS,Walti}.
For superconductors with gap nodes, PCS can in general provide
information on the pairing symmetry \cite{Alff,Weiprl,Mao,Biswas}.
In this Letter, we report PCS measurements on single crystals of
CeCoIn$_5$ in the temperature range 150 mK to 2.5 K. We observed
Andreev-reflection characteristics with multiple structures, whose
dependence on junction impedance indicates two \emph{coexisting}
order parameter components (OP) with nodal characteristics. These
OP's show sizable amplitudes relative to $T_c$ and different
evolutions with temperature. Our observations suggest a highly
unconventional pairing mechanism, possibly involving multiple bands.

In PCS, electronic transmission between a normal metal and a
superconductor is measured as conductance $dI/dV$ versus bias
voltage $V$ across a ballistic contact junction. For a transparent
contact, $dI/dV$ is primarily determined by Andreev reflection,
based on the conversion of electrons or holes into Cooper pairs,
which doubles $dI/dV$ inside the superconducting energy gap. For
non-transparent junctions, $dI/dV$ involves both Andreev reflection
and quasiparticle tunneling. The standard model for calculating
$dI/dV$ was given by the Blonder-Tinkham-Klapwijk (BTK) theory for
$s$-wave pairing~\cite{BTK}, and subsequently generalized for
$d$-wave pairing~\cite{Hu,TK}. A key spectral signature of the
$d$-wave gap nodes is the zero-bias conductance peak, which arises
from $\emph{surface}$ states bound by phase interference between
consecutively Andreev-reflected quasiparticles~\cite{Aprili}. This
peak structure is to be distinguished from the hump structure
associated with conventional Andreev \emph{bulk} states. In the
generalized BTK scenario, relative manifestation of the Andreev
surface versus bulk states depends on both junction orientation and
a dimensionless parameter $Z$ representing junction impedance
\cite{Z-parameter}, thus allowing the OP to be studied
~\cite{TK,Weiprl}.

The single crystals of CeCoIn$_5$ used in this work were grown by a
self-flux method~\cite{Petrovic}, and characterized by both x-ray
diffraction and magnetic susceptibility to confirm material
uniformity.  The crystals were platelets approximately
1$\times$1$\times$0.2 mm$^3$ in size, each showing a sharp
superconducting transition at $T_c=2.3$ K.  The crystal surfaces
were etched with HCl and rinsed with ethanol prior to measurement,
in order to remove any residual In flux. High purity Pt-Ir tips were
used as normal-metal electrodes, gently pressed onto the $c$-axis
face of each crystal with a spring-cushioned differential
micrometer. This point-contact mechanism was attached to the mixing
chamber of a high cooling-power $^3$He/$^4$He dilution refrigerator,
and enabled the junction impedance to be varied \emph{in situ} at
low temperatures.  The point contacts we measured were in the 0.2-1
$\Omega$ range, consistent with the contact size being in the
ballistic regime~\cite{Sharvin}. To minimize Joule heating in the
junction over a wide voltage range, our spectroscopy data was
acquired by a pulsed technique: 2 ms current pulses were applied
through the contact in 20\% duty cycles, and the junction voltage
was measured 80 times within each pulse and then averaged. The $I$
vs $V$ curves were obtained by varying the current level, and then
numerically differentiated to obtain the $dI/dV$ vs $V$ spectra.

\begin{figure}
 \centering
 \includegraphics[width=2.4in]{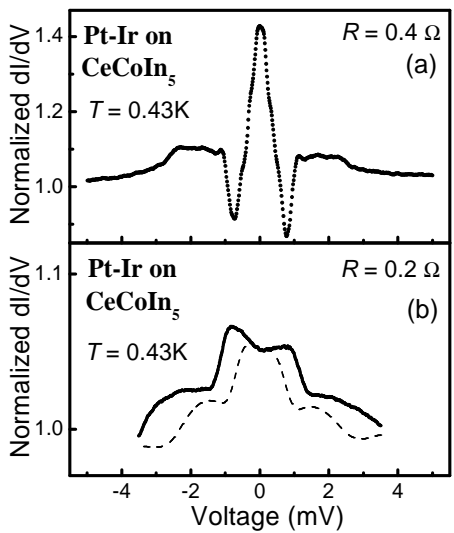}
 \caption{\label{fig:data}
 Normalized $dI/dV$ vs $V$ data for Pt-Ir point contacts on
CeCoIn$_5$ at 0.43 K. Top panel (a) is for a 0.4 $\Omega$ junction.
Bottom panel (b) is for a 0.2 $\Omega$ junction, with the 1.5 K
curve (dashed) also plotted to clearly show the double humps.}
\end{figure}

Two types of spectra were observed in our measurements, depending on
the point-contact impedance.  These measurements were reproducible
on multiple spots over different samples and repeated at each spot
to rule out any surface destruction by the point contact.
Figure~\ref{fig:data} shows $dI/dV$ spectra taken at 0.43 K well
below $T_c$, after normalization relative to spectra taken above
$T_c$. The top panel is for a 0.4 $\Omega$ junction, and the bottom
panel is for a 0.2 $\Omega$ junction. Distinct spectral features are
seen in the top panel, with a sharp zero-bias peak dipping at $\sim$
$\pm$ 1 mV into a broad spectral hump $\sim$ $\pm$ 2.5 mV in width.
Small kinks are also visible on the peak at $\sim$ $\pm$ 0.3 mV and
$\sim$ $\pm$ 0.5 mV.  The main peak, dip and hump structures evolve
differently with decreasing junction impedance. As seen in the
bottom panel, the peak becomes an asymmetric inner hump $\sim$ $\pm$
1 mV in width, the dips get filled in, while the outer hump remains
largely unchanged. These hump structures are the classic signatures
of Andreev reflection, which introduces excess spectral states
inside the energy gap \cite{BTK}.  These excess states expectedly
diminish with temperature, as is evident in the 1.5K data [dashed
curve in Fig.1(b)].  The zero-bias peak, on the other hand, is key
evidence for nodes in the gap \cite{Hu}. It is worth noting that
peak and hump structures of similar shapes and energy scales have
been reported in an earlier PCS study of CeCoIn$_5$, although
appearing separately in different spectra~\cite{Goll}. Our measured
spectra are clearly hybrid in character, each containing multiple
structures.

\begin{figure}
 \centering
 \includegraphics[width=2.4in]{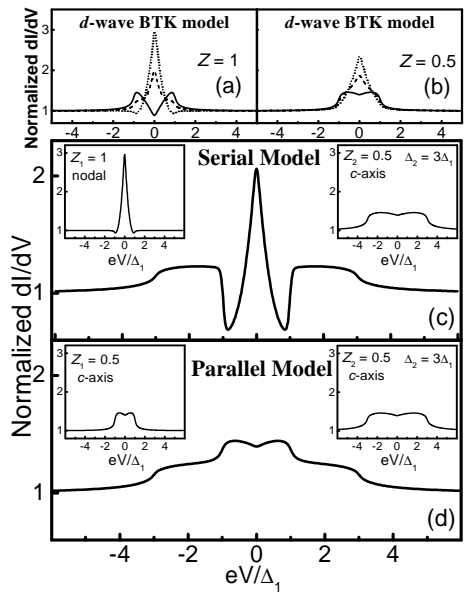}
 \caption{\label{fig:simulations}
Spectral simulations using the generalized $d$-wave BTK formalism.
 (a) and (b) are for $Z = 1$ and $Z = 0.5$ junctions, with nodal
(dotted curves), antinodal or $c$-axis (solid curves), and
angle-averaged (dashed curves) orientations.  (c) and (d) show
serial and parallel superpositions of two spectral contributions
(insets) for two OP's with $\Delta_2$=3$\Delta_1$ and different
$Z$'s.}
\end{figure}

To identify the multiple spectral features observed in our data, we
consider theoretical spectra from the generalized BTK model. Shown
in the top panels of Fig.~\ref{fig:simulations} are the simulated
$dI/dV$ spectra for a $d$-wave OP, plotted in normalized units vs
$eV/\Delta$, where $\Delta$ is the $d$-wave gap
maximum~\cite{temperature}. The choice of $d$-wave symmetry here is
motivated by both thermodynamic and transport
data~\cite{Izawa-aniz,Aoki}, and intended to illustrate the generic
spectral dependence on junction orientation and impedance.  The
curves in Fig. 2(a) are for a high-impedance ($Z = 1$) junction, and
the curves in Fig. 2(b) are for a low-impedance ($Z = 0.5$)
junction. In each plot, the dotted/solid curve is for a
nodal/antinodal junction (normal to a nodal/antinodal axis), while
the dashed curve models the effect of junction roughness by
averaging over all intermediate orientations. Note that an ideal
$c$-axis junction would produce similar spectra as the antinodal
case, since there is no OP sign change about the junction normal in
either case to allow for Andreev interference. The overall spectral
evolution between peak and hump structures is a direct manifestation
of the competition between Andreev surface and bulk
states~\cite{Hu}.

From these generic spectral simulations, the data in
Fig.~\ref{fig:data} can be interpreted as the \emph{superposition}
of two types of spectral contributions.  Namely, the sharp peak
structure comes from Andreev surface states due to a high-$Z$ nodal
junction, and the broad hump structures come from Andreev bulk
states due to low-$Z$ antinodal junctions. The appearance of two
effective $Z$'s, with very different dependences on junction
impedance, is indicative of different Andreev coupling to two
distinct OP's. To demonstrate this two-OP scenario, we have
developed a superposition model, based on the ``serial" precedence
of surface over bulk states in junction transmission. More
specifically, when bulk spectra from two different OP's coexist,
their superposition is essentially $\emph{additive}$~\cite{Szabo},
since bulk states can be accessed in ``parallel". However, when
\emph{both} surface and bulk spectra are involved, the junction
transmission becomes effectively ``serial", thus justifying a
$\emph{multiplicative}$ superposition within energies ($|eV| <
\Delta$) where Andreev surface states can readily form. This serial
model is demonstrated in Fig.~\ref{fig:simulations}(c), by
superposing a peak spectrum (left inset) with a hump spectrum (right
inset) of triple the energy scale (i.e. $\Delta_2 = 3\Delta_1$).
Here the component spectra were multiplied for $|eV| < \Delta_1$ and
added for $|eV| > \Delta_1$, following our model justifications. The
peak-dip-hump structures seen in the data of Fig.~\ref{fig:data}(a)
are remarkably well reproduced here in
Fig.~\ref{fig:simulations}(c).  For comparison, the parallel model
superposing two low-$Z$ bulk spectra (insets) is shown in Fig. 2(d),
also generically reproducing the multiple-hump data seen in
Fig.~\ref{fig:data}(b). The overall spectral resemblance between our
simulations and data is robust evidence for the coexistence of two
OP's.

Some general remarks about our two-OP spectral analysis should be
made. First, our model was intended to show generically how two
coexisting OP's with gap nodes could produce the multiple spectral
structures observed. The distinctively serial relationship between
the peak and hump structures clearly establishes the surface-state
nature of the former, as arising from Andreev interference for a
nodal OP. However, although our data can be explained within a
$d$-wave framework, we cannot rule out the presence of other OP line
or point nodes, along either the pole or the equator, such as in the
case of UPt$_3$~\cite{Joynt}. Precise determination of the pairing
symmetry in CeCoIn$_5$ would require a systematic study of the
spectral anisotropy~\cite{Weiprl}, along with an extension of the
generalized BTK theory beyond its two-dimensional formulation.
Second, the non-trivial spectral evolution we observed versus
junction impedance indicates a complex $k$-space dependence of $Z$,
with the nodal-junction states dominating at high $Z$ and
antinodal-junction states dominating at low $Z$. While surface
roughness could allow for nodal-junction surfaces to exist on a
nominally $c$-axis crystal, a detailed explanation of the
peak-to-hump evolution would require full understanding of how $Z$
depends on the complex band structure of CeCoIn$_5$
\cite{Z-parameter}.  For example, multiband coupling could in theory
affect the formation of Andreev surface states \cite{VolkerSigrist}.
The effects of band structure on quasiparticle tunneling have also
been studied \cite{Weiprb}. Third, the spectral heights tend to be
smaller in the data than in the model, a difference which could be
attributed to non-superconducting spectral contributions from either
uncondensed quasiparticles \cite{Agterberg} or Kondo scattering
\cite{Fisk}.

\begin{figure}
 \centering
 \includegraphics[width=2.4in]{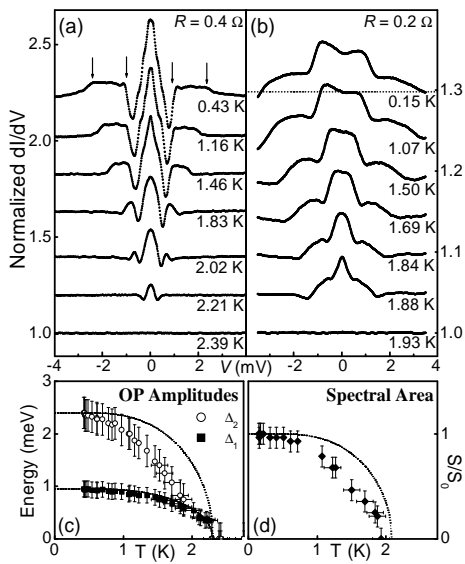}
 \caption{\label{fig:tdep}
Temperature dependence of the data from Fig.~\ref{fig:data}. A
subset of the spectral evolutions are shown in (a) and (b). The OP
amplitudes $\Delta_1(T)$ and $\Delta_2(T)$ determined from (a) are
plotted in (c).  The reduced spectral area $S/S_0$ extracted from
(b) is plotted in (d).  Theoretical BCS curves (dotted) are included
to indicate deviations from mean-field behavior.}
\end{figure}

The temperature dependence of our spectral data was also examined.
Figure~\ref{fig:tdep} shows spectral evolution of the data from
Fig.~\ref{fig:data} in the temperature range 150 mK to 2.5 K. The
spectra are staggered for clarity, with arrows in Fig.~3(a) to
indicate the two-OP amplitudes determined from the serial model
above, and a dotted baseline in Fig.~3(b) to underscore the
``excess" spectral area associated with bulk Andreev states.  The OP
amplitudes $\Delta_1$($T$) and $\Delta_2$($T$) are plotted in
Fig.~\ref{fig:tdep}(c), along with theoretical (dotted) curves
calculated from the BCS gap equation. The excess spectral area $S$
is similarly plotted in Fig.~\ref{fig:tdep}(d), after normalization
by its base-temperature value $S_0$ \cite{DeWilde}. From Fig.~3(c),
it is clear that both OP amplitudes approach distinct
zero-temperature values, $\Delta_1 = 0.95 \pm 0.15$ meV and
$\Delta_2 = 2.4 \pm 0.3$ meV, and vanish near $T_c = 2.3$ K,
consistent with both OP's being components of the same
superconducting order. This common $T_c$ also argues against the
presence of a proximity-induced superconducting layer in our
junctions, which should cause the smaller order parameter component
to vanish below the bulk $T_c$ \cite{Szabo}. However, while
$\Delta_1$($T$) is well described by the BCS gap equation,
$\Delta_2$($T$) deviates markedly from mean-field behavior.  This
deviation is also evident in the reduced spectral area ($S/S_0$)
plot in Fig.~3(d), indicating a predominance of the larger OP for
parallel superposition. Similar deviations have been observed in
other heavy-fermion superconductors, and attributed to the nodality
of highly complex pairing symmetries~\cite{DeWilde,HFS,Walti}.
Alternatively, the difference between $\Delta_1$($T$) and
$\Delta_2$($T$) could be the signature of novel interplay between
two different types of order~\cite{Mathur,Demler}.

Finally we discuss the physical implications of our results on the
pairing mechanism in CeCoIn$_5$.  First, assuming that each of the
two energy scales identified above can be directly assigned to a
superconducting OP, they would correspond to gap-to-$T_{c}$ ratios
of $2\Delta_1/k_{B}T_{c} = 9.5 \pm 1.5$ and $2\Delta_2/k_{B}T_{c} =
24 \pm 3$. These ratios are much larger than the BCS weak-coupling
value of 3.5 for phonon-mediated pairing, and well beyond the
strong-coupling limit even after $d$-wave
corrections~\cite{Carbotte}. One conceivable way to enhance the
gap-to-$T_{c}$ ratio is through inter-band coupling, whereby
carriers from different bands could interact to result in multiple
pair potentials sharing a common $T_c$~\cite{Suhl}. This multi-band
scenario would be physically plausible for CeCoIn$_5$, considering
that its Fermi surface has four distinct sheets with different
topologies and effective masses~\cite{Bandstructure,dHvA}.
Furthermore, Andreev scattering for a heavy-mass 2D sheet would be
inherently weaker than for a light-mass 3D sheet, due to poorer
Fermi-velocity matching across the junction~\cite{Z-parameter}. This
multi-band effect could provide a natural explanation for the two
different $Z$ scales observed in our spectra. However, even allowing
for inter-band coupling between highly disparate densities of
states~\cite{Suhl}, a sizable ``intrinsic" $2\Delta/k_{B}T_{c}$,
intermediate between $\approx$ 9.5 and 24, may still be needed to
explain our data~\cite{Suhl,Iavarone}.  Such an intrinsically large
gap-to-$T_{c}$ ratio would present a serious challenge to current
theoretical formulations~\cite{Carbotte,BangBalatsky}, at least
within the Fermi-liquid framework, thus indicating a highly
unconventional pairing mechanism in CeCoIn$_5$.

In summary, we have performed point-contact spectroscopy on the
heavy-fermion superconductor CeCoIn$_5$.  Andreev-reflection
characteristics with multiple structures were observed.  Spectral
analysis using the generalized BTK formalism revealed two coexisting
order parameter components with nodal symmetry and sizable
amplitudes. These observations suggest a highly unconventional
pairing mechanism in a multi-band scenario.

Work supported by: NSERC, CFI/OIT, Canadian Inst. for Advanced
Research; Division of Materials Sciences, Office of Basic Energy
Sciences, US Dept. of Energy under Contract No. DE-AC02-98CH10886.


\begin{references}
\bibitem{Sharvin} Using the Sharvin formula for the contact resistance $R=$~4$\rho l / 3 \pi a^2$, with $\rho \sim
~3 \mu\Omega$cm \cite{Petrovic} and $l \sim 80$ nm at $T_c$
\cite{Movshovich-unconv} for CeCoIn$_5$, we estimate our contact
radius $a < 70$ nm, satisfying the ballistic criterion $a < l$
\cite{BTK}. This criterion would be further justified if a
lower-temperature $\rho \sim 0.3~\mu\Omega$cm \cite{Paglione} were
used, giving $l \sim 800$ nm.
\bibitem {temperature} In our simulations, temperature was fixed at $k_{B}T = 0.03\Delta$, consistent with experimental
conditions.
\end{references}
\end{document}